\documentstyle[12pt,psfig]{article}

\catcode`\@=11
\def\marginnote#1{}
\newcount\hour
\newcount\minute
\newtoks\amorpm
\hour=\time\divide\hour by60
\minute=\time{\multiply\hour by60 \global\advance\minute by-
\hour}
\edef\standardtime{{\ifnum\hour<12 \global\amorpm={am}%
    \else\global\amorpm={pm}\advance\hour by-12 \fi
    \ifnum\hour=0 \hour=12 \fi
    \number\hour:\ifnum\minute<100\fi\number\minute\the\amorpm}}
\edef\militarytime{\number\hour:\ifnum\minute<100\fi\number\minute}
\def\draftlabel#1{{\@bsphack\if@filesw {\let\thepage\relax
  \xdef\@gtempa{\write\@auxout{\string
    \newlabel{#1}{{\@currentlabel}{\thepage}}}}}\@gtempa
    \if@nobreak \ifvmode\nobreak\fi\fi\fi\@esphack}
     \gdef\@eqnlabel{#1}}
\def\@eqnlabel{}
\def\@vacuum{}
\def\draftmarginnote#1{\marginpar{\raggedright\scriptsize\tt#1}}
\def\draft{\oddsidemargin -.5truein
        \def\@oddfoot{\sl preliminary draft \hfil
        \rm\thepage\hfil\sl\today\quad\militarytime}
        \let\@evenfoot\@oddfoot \overfullrule 3pt
        \let\label=\draftlabel
        \let\marginnote=\draftmarginnote
   
\def\@eqnnum{(\theequation)\rlap{\kern\marginparsep\tt\@eqnlabel}%
\global\let\@eqnlabel\@vacuum}  }
\def\preprint{\twocolumn\sloppy\flushbottom\parindent 1em
        \leftmargini 2em\leftmarginv .5em\leftmarginvi .5em
        \oddsidemargin -.5in    \evensidemargin -.5in
        \columnsep 15mm \footheight 0pt
        \textwidth 250mmin      \topmargin  -.4in
        \headheight 12pt \topskip .4in
        \textheight 175mm
        \footskip 0pt
        
\def\@oddhead{\thepage\hfil\addtocounter{page}{1}\thepage}
        \let\@evenhead\@oddhead \def\@oddfoot{} \def\@evenfoot{} 
}
\def\titlepage{\@restonecolfalse\if@twocolumn\@restonecoltrue\onecolumn
     \else \newpage \fi \thispagestyle{empty}\c@page\z@
        \def\thefootnote{\fnsymbol{footnote}} }
\def\endtitlepage{\if@restonecol\twocolumn \else  \fi
        \def\thefootnote{\arabic{footnote}}
        \setcounter{footnote}{0}}  %\c@footnote\z@ }
\catcode`@=12
\relax
\def\be{\begin{equation}}
\def\ee{\end{equation}}
\def\bea{\begin{eqnarray}}
\def\eea{\end{eqnarray}}
\def\simlt{\stackrel{<}{{}_\sim}}
\def\simgt{\stackrel{>}{{}_\sim}}
\def\NPB#1#2#3{{\it Nucl.~Phys.} {\bf B#1} (19#2) #3}
\def\PLB#1#2#3{{\it Phys.~Lett.} {\bf #1B} (19#2) #3}
\def\PRD#1#2#3{{\it Phys.~Rev.} {\bf D#1} (19#2) #3}
\def\PRL#1#2#3{{\it Phys.~Rev.~Lett.} {\bf #1} (19#2) #3}
\def\ZPC#1#2#3{{\it Z.~Phys.} {\bf C#1} (19#2) #3}

\def\PR#1#2#3{{\it Phys.~Rep.} {\bf #1} (19#2) #3}

\relax
\begin{document}
%\draft
%\preprint
%
\begin{titlepage}
\begin{flushright}
CERN-TH/95-286\\
ROME1-1121/95\\
hep-ph/9511259 \\
\end{flushright}
\vskip 0.3in
\begin{center}{\Large\bf 
Light Higgsino Detection at LEP1.5} \\
\vskip .7in
{\bf S. Ambrosanio$^*$, $\;$ B. Mele} \\
~\\
{\it Dipartimento di Fisica, Universit\`a \lq\lq La Sapienza'' 
and I.N.F.N., Sezione di Roma, \\
P.le Aldo Moro 2, I-00185 Rome, Italy}\\
%e-mail: {\tt ambrosanio@roma1.infn.it, mele@roma1.infn.it}  
~\\
~\\ 
{\bf M. Carena, $\;$ C.E.M. Wagner}\\
~\\
{\it CERN, TH Division, CH--1211 Geneva 23, Switzerland}\\
%e-mail: 
\end{center}
\vskip 1.2cm
\begin{center}
{\bf Abstract}
\end{center}
\begin{quote}
Within the minimal supersymmetric 
extension of the Standard Model,
the best fit to the most recent precision-measurement data 
requires charginos and  neutralinos, with 
dominant Higgsino components and with masses within the reach of
LEP1.5 ($\sqrt{s}=140$ GeV). In this work, we present
a detailed analysis of the neutralino and chargino production 
processes for the favoured region of parameter space, that is
low values of $|\mu|$ and either low or 
large values of $\tan\beta$. We find that  chargino and
neutralino searches can cover the Higgsino region in the
($\mu,M_2$) plane for values of $M_2 \simlt 1$~TeV,
at the next phases of the LEP collider. We also 
show that, due mainly to phase-space constraints, 
the lightest neutralinos    should be
more easily detectable than charginos 
in most of the parameter space preferred by precision-measurement
data.
\end{quote}
\vskip1.5cm

\begin{flushleft}
CERN-TH/95-286 \\
October 1995 \\
\end{flushleft}

\vspace*{\fill}

\noindent 
\parbox{0.4\textwidth}{\hrule \hfill}  

\noindent 
{\small 
$^*$\   Address after December 1, 1995: University of Michigan, 
        3068A H.M. Randall Physics Lab.,\\ 
        $\phantom{^*\;}$ 500 East University, 
        Ann Arbor, MI 48109-1120, U.S.A.} 

\end{titlepage}
\newpage
\setcounter{footnote}{0}
\setcounter{page}{0}
\thispagestyle{empty} 
\null
\newpage
%
% BODY
\noindent
The minimal supersymmetric extension of the Standard Model 
allows a solution of the gauge hierarchy problem and can
be obtained as a low energy effective theory of 
   supersymmetric grand unified
theories including also gravity \cite{SUSYG}. The model contains
a large number of free parameters associated with the soft breakdown
of supersymmetry, which lead to large uncertainties in the
supersymmetric particle
discovery potential of present and future colliders \cite{hk}.
In particular, the physical properties of the chargino and
neutralino sectors of the theory 
depend on four unknown parameters \cite{neumatr}:
the soft supersymmetry breaking mass parameters of the supersymmetric
partners of the  $SU(2)_L$ and $U(1)_Y$ gauge bosons, which we
shall denote by $M_2$ and $M_1$, respectively,
the Higgs supersymmetric mass parameter $\mu$, and the ratio of Higgs
vacuum expectation values $\tan\beta$. If the theory proceeds 
from a grand unified theory at very high energies, which will be implicitly
assumed within this work, the low energy
values of the gaugino masses  are related by 
$M_1 \simeq M_2 \alpha_1/\alpha_2$, reducing to three 
the number of relevant free parameters.

The existence of light supersymmetric particles in nature can be tested
through  direct experimental detection,    and also through deviations of
the precision measurement data from the Standard Model predictions.
Supersymmetric particles can, indeed, affect the low-energy observables
through    loop radiative corrections, which become negligibly small
as soon as the supersymmetric particle masses are far above the electroweak
scale. The most recent precision-measurement data show that the
ratio of the width of $Z$ to bottom quarks
to its total hadronic decay width, $R_b = \Gamma_b/\Gamma_h$, is
more than three standard deviations above the Standard Model
prediction for this quantity \cite{lepglobal}.  
This deviation can be partially explained by
the presence of  supersymmetric particles at the 
weak scale. The best fit to the precision-measurement data leads 
to strong  constraints  on  two
of the three independent parameters of the neutralino and chargino 
sector of the theory: $\mu$ and $\tan\beta$ \cite{BF}--\cite{Hollik}. 

Within the MSSM, large one-loop corrections to $R_b$ 
are always associated
with large values of the third-generation Yukawa couplings. The 
supersymmetric top 
and bottom Yukawa couplings are related to their running mass values by: 
\be 
h_t = \frac{m_t}{\sin\beta}, \;\;\;\;\;\;\;\;\;\;\;
h_b = \frac{m_b}{\cos\beta}.
\label{topbot}
\ee 
From Eq. (\ref{topbot}), it is clear that in order to enhance the
top (bottom) quark Yukawa coupling effects, $\tan\beta$ should acquire the 
smallest (largest) value allowed by the theory \cite{Gordy,Stefan}.
The requirement that the Yukawa couplings remain in the perturbative
domain    up to  scales of the order of the grand unification scale 
$M_{GUT} \simeq 10^{16}$ GeV
implies that the largest corrections are obtained either at values
of $\tan\beta \simeq m_t/m_b$  \cite{Joan} or at the infrared 
fixed-point solution for the top quark mass \cite{CW}. 
Interestingly enough, for the values of the top quark mass 
measured at the Tevatron collider \cite{topquark}, 
these regions of the parameter
space are also preferred \cite{talk} from the point of view 
of Yukawa coupling unification 
at a scale $M_{GUT}$ \cite{Yukawa,largetb}.  

In the low $\tan\beta$ regime, large positive corrections to $R_b$
may only be obtained through the one-loop 
chargino-stop contributions \cite{BF},
which are enhanced for light stops, predominantly right-handed, 
and for chargino masses close to $M_Z/2$. 
These effects are most relevant for   
low $|\mu|$ values, i.e.  $|\mu| \simlt M_Z$, 
for which the Higgsino component of the 
lightest chargino is enhanced and the chargino-stop-bottom coupling
is approximately given by $h_t$, Eq. (\ref{topbot}). For large values 
of $\tan\beta$, large positive corrections may also be obtained 
from both the neutral
Higgs sector of the theory and from neutralino-sbottom loops.
Similarly to the low 
$\tan\beta$ case, the genuine supersymmetric contributions
are maximized for low values of the supersymmetric mass 
parameter $|\mu| \simlt M_Z$.
There are also strong restrictions on the lightest stop and sbottom
particles, which should be light and predominantly 
right-handed to avoid unacceptable corrections to the $W^\pm$ mass
or to the $Z$ leptonic width, but do not play a direct role 
in the chargino and  neutralino production processes.

In this paper, we study the potential of LEP1.5 
(the LEP phase at $\sqrt{s}\simeq140$ GeV that
started running
at the end of October 1995)
to explore the range of parameters 
suggested by precision measurements
through  neutralino and chargino searches. 
The phase 1.5 of the LEP collider is expected to collect
an integrated luminosity of 5--10~${\rm pb}^{-1}$.
We also extend our analysis to the higher-energy 
($\sqrt{s}\simeq190$ GeV) operation phase of LEP2.

Neutralino production at LEP can proceed 
through the process  \cite{bartl}--\cite{amb-mele}: 
\be
e^+ e^- \rightarrow \widetilde{\chi}^0_1 \widetilde{\chi}^0_2 \; , 
\label{process}
\ee 
where $\widetilde{\chi}^0_1$ and $\widetilde{\chi}^0_2$ are the
lightest and next-to-lightest neutralino particles, respectively.
This leads to a very interesting signal, since the 
$\widetilde{\chi}^0_1$ escapes detection and the $\widetilde{\chi}^0_2$
decay products are hence completely unbalanced in energy and momentum.
Pair production of the lightest neutralino contributes to
the $Z$-boson invisible decay width and is hence not relevant
for neutralino detection at energies above the $Z$ pole.

Light charginos are mainly produced in pairs \cite{Tata,Bartl-ch}: 
\be
e^+ e^- \rightarrow \widetilde{\chi}^+_1 \widetilde{\chi}^-_1
\ee 
and, due to their short lifetime, 
can be searched for through their three-body decays into the lightest
neutralino plus a light fermion pair.

For particular parameter configurations, 
the production rate of light charginos and neutralinos at LEP   
also suffers from large uncertainties related to the neutral
and charged slepton spectrum, respectively. 
The uncertainties are  larger  
when the gaugino components of the neutralinos/charginos are more 
important. Indeed, due to the smallness of the electron Yukawa coupling,  
the strength of the slepton coupling to electrons is just a reflection of 
the largeness of the gaugino components in charginos and neutralinos. 
On the other hand,
if the light charginos and neutralinos are predominantly Higgsinos, 
as suggested by precision measurements, the main production mechanism
is through {\it s}-channel 
$Z/\gamma$ exchange and the production cross sections 
may be accurately determined. Requiring chargino masses smaller than
$M_Z$, Higgsino dominance occurs for
values of $|\mu| \simlt M_Z$. In this work, we will concentrate on the range 
$2 M_Z < M_2 \simlt 1$~TeV and $|\mu| \simlt M_Z$,  
in which the ratio $M_2/|\mu|$ guarantees
a dominant Higgsino component in light neutralinos/charginos  
and the values of $M_2$  can still be considered ``natural".

As we will show, the production rates are large for both neutralinos 
and charginos in the above scenario. On the other hand, when 
$M_2$ is increased,
$\widetilde{\chi}^+_1$ and $\widetilde{\chi}^0_1$ tend 
to be degenerate in mass.
The same holds, but to a lesser extent, for ${\widetilde\chi}^0_2$ and 
$\widetilde{\chi}^0_1$. Indeed, the relation  
$m_{\widetilde{\chi}^0_2}   > 
 m_{\widetilde{\chi}^\pm_1} > 
 m_{\widetilde{\chi}^0_1}$ holds in  all 
the regions of the plane ($\mu, M_2$) 
considered here (apart from a small region already excluded 
by LEP1 data).
The actual production of 
$\widetilde{\chi}^0_1 \widetilde{\chi}^0_2$ and
$\widetilde{\chi}^+_1 \widetilde{\chi}^-_1$ pairs is observed through
the decays \footnote{Additional missing energy,
in the form of neutrinos, may be present}: 
\bea
\widetilde{\chi}^0_2  &  \rightarrow &  \widetilde{\chi}^0_1 
+ \; {\rm (visible)} ; \\
\widetilde{\chi}^{\pm}_1 &  \rightarrow & \widetilde{\chi}^0_1 + 
{\rm (visible)}. 
\eea

Since $\widetilde{\chi}_2^0$ ($\widetilde{\chi}^\pm_1$)
is produced with rather low velocity,
if the mass splitting between the next-to-lightest
neutralino (chargino) and the lightest neutralino is small, 
the decay products may not
have sufficient energy and multiplicity to pass the normal
experimental trigger for missing energy and momentum events.
Indeed, in the Higgsino region, these mass splittings are 
naturally small, of order $M_W^2/M_2$.
Presently, in Monte Carlo simulations that study supersymmetric signals and 
backgrounds at LEP \cite{interim}, a mass splitting:  
\bea 
\Delta_{\pm} & \equiv & m_{\widetilde{\chi}^\pm_1} - 
m_{\widetilde{\chi}^0_1}  \simgt 10 \; {\rm GeV} \; , \label{phase1} \\ 
\Delta_{0}   & \equiv & m_{\widetilde{\chi}^0_2} - 
m_{\widetilde{\chi}^0_1}  \simgt 10 \; {\rm GeV}\; , \label{phase2}
\eea 
is required for the events to be observable.
In the following, we prove that, after imposing
cuts on the mass splittings as in Eqs. (\ref{phase1}) and 
(\ref{phase2}),
neutralino searches are more efficient than chargino searches
in covering the ($\mu,M_2$) Higgsino region. 

Cross sections for the processes  Eqs.(2) and (3) at the LEP1.5 energy
are shown in Figs. 1 and 2, at low values (close to the
fixed point value for $M_t = 170$ GeV) and large values 
of $\tan\beta$,  as preferred by precision measurements, 
in the plane ($\mu,M_2$).
In particular, we have fixed $\tan\beta = 1.2$ and 50, respectively.
%%%%%%%% NEW  %%%%%%%%%
The effect of QED initial state radiation is included.
All the slepton masses are set  at 300 GeV.
%%%%%%%% NEW  %%%%%%%%%
However, we checked that varying slepton masses between 50 GeV and 1 TeV
does not cause any noticeable change in the cross sections, 
in the range of ($\mu,M_2$) considered here.
We have also checked that the low $\tan\beta$ results are basically
unchanged for $\tan\beta$ values that are closer to (or somewhat 
further away from) 1. 

The LEP1 and LEP1.5 kinematical limits are shown in the figures.
One can observe that, in almost all the region 
kinematically covered by LEP1.5, 
cross sections are as large as several picobarns 
for both low and large values of $\tan\beta$.
For an integrated luminosity of 10 ${\rm pb}^{-1}$, one expects
%%%%%%%% NEW  %%%%%%%%%
up to more than 100  events from neutralino
%%%%%%%% NEW  %%%%%%%%%
(chargino) production,  before experimental cuts are applied.

In all the figures, dashed lines give contours for 
the splitting of the decaying particle mass
and the lightest neutralino mass, $\Delta_0$ and $\Delta_{\pm}$.
One can see that, in general, imposing the conditions 
in Eqs. (\ref{phase1}) and (\ref{phase2}) excludes a much larger
portion of the parameter space in the chargino case, for both
low and high $\tan\beta$ values.
On the other hand,
for neutralinos the phase-space restriction 
in Eq. (\ref{phase2}) is met
up to very large values of $M_2$, of order 1 TeV. 
While chargino cross sections are comparable to the neutralino 
ones, $\widetilde{\chi}^{\pm}_1$ is always considerably closer 
in mass to $\widetilde{\chi}^0_1$ than  
$\widetilde{\chi}^0_2$, so that
chargino detection becomes more difficult at high  $M_2$, mostly due
to phase-space constraints. 
In particular, Eq. (\ref{phase1}) is fulfilled only for
$M_2 \simlt (5$--$6) M_Z$ in both the low and high 
$\tan\beta$  regime. Moreover, the larger unbalance in energy
and momentum in the case of neutralino final states further
enhances the advantages of the $\widetilde{\chi}_1^0
\widetilde{\chi}_2^0$ channel.

Hence, if light neutralinos with a mass below the kinematic limit and 
a dominant Higgsino component are present in the theory, they should be 
more easily detected, at the next phase of the LEP collider.
Possible invisible decays
of the next-to-lightest neutralino and cascade decays through
a chargino (that are even more constrained by phase space)
can deplete the observable neutralino production rate. On the other
hand, one can see that, for Higgsino dominated 
compositions,  the $\widetilde{\chi}^0_2$ 
decays into neutrinos with a BR less than 20\% and the cascade-decay 
fraction can reach at most 20--30\%~\cite{amb-mele2}. 
Hence, most of the $\widetilde{\chi}^0_2$ decays
should have three body visible final states. 

We stress that
the mass splitting we are requiring for neutralino (chargino)
detection, Eqs. (\ref{phase1}) and (\ref{phase2}), is based on the
present experimental analysis \cite{dionisi}, which shows that  
neutralino (chargino) searches for mass splittings
5 GeV$<\Delta_0(\Delta_{\pm})<10$ GeV are remarkably more difficult
and require a dedicated experimental trigger on the energy and 
multiplicity of the events.
More extreme cases with $\Delta_0(\Delta_{\pm})<5$ GeV
seem, of course, even more challenging.
It is clear, however, that a definite  statement 
on the neutralino and chargino observability
requires a detailed study
of the signal versus possible backgrounds through Monte Carlo
simulations that can mimic the effect of experimental
triggers and kinematical cuts.

We have also considered Higgsino production at LEP2 (Figs. 3 and 4).
The last phase of the LEP collider will reach an energy of about
190 GeV and a total luminosity of 300 ${\rm pb}^{-1}$. In this case,
although the total cross section goes down
by a factor 3 or 4, the luminosity is
sufficiently high  to discriminate the presence of light
supersymmetric particles in  a clear way. However,  
the production processes at these energies 
 suffer from a large background from $W^+W^-$ and $ZZ$ production 
in both the hadronic and leptonic modes, 
which is suppressed at the LEP1.5 center of mass energies.
One can see in Figs. 3 and 4 that the outcome at LEP2 is analogous 
to the one at LEP1.5. The high-$M_2$ region is 
more challenging for chargino than for neutralino searches.
Nevertheless, despite the higher luminosity,
the background analysis could be more involved at larger~$\sqrt{s}$.

In conclusion, we have shown that there is a clear indication that
neutralino searches are at least as competitive as chargino ones, 
to explore  the Higgsino region in the ($\mu,M_2$) plane. 
In particular, neutralino  searches can cover 
this region of parameters up to almost 
the kinematical limit, 
for values of $\tan\beta$ favoured by precision measurements
and values of $M_2 \simlt 1$ TeV. In the case of charginos, the
kinematic reach in the ($\mu,M_2$) plane is almost equivalent.
On the other hand, the phase-space constraints in chargino decays 
make its detection more challenging than in  neutralino production
processes. A similar result is valid also for intermediate 
$\tan\beta$ values. \\
~\\
~\\
{\bf Acknowledgements:} We would like to thank P.~Chankowski, C.~Dionisi, 
M.~Felcini and J.~F.~Grivaz for useful comments and discussions.
%%%%%%%% NEW  %%%%%%%%%
Research supported in part by the European Union under contract
No.~ERB-CHRX-CT93-0132.
%%%%%%%% NEW  %%%%%%%%%
\newpage
%%%%%%%%%%%%%%%%%%%%%%%%%%%%%%%%%%%%%%%%%%%%%%%%%%%%%%%%%%%%

\newpage 

\noindent 
{\Large \bf Figure captions} 

\vskip 1.0cm

\begin{description} 

\item[Fig.~1] Cross section and kinematics, in the $(\mu, M_2)$ plane, 
of the processes 
\mbox{$e^+ e^- \rightarrow \widetilde{\chi}^0_1 \widetilde{\chi}^0_2$} (a) 
and 
\mbox{$e^+ e^- \rightarrow \widetilde{\chi}^+_1 \widetilde{\chi}^-_1$} (b) 
at LEP1.5, in the $\tan\beta = 1.2$ case. 
Contour lines for cross sections (in pb) are represented by solid black lines. 
The grey lines represent the LEP1 and LEP1.5 kinematical reach for the 
process. 
The dashed lines and the bold labels in Fig. 1a (b) give isocontours 
in GeV for the quantity $\Delta_0$ ($\Delta_\pm$), defined in 
Eq. (\ref{phase1}) (Eq. (\ref{phase2})). 

\item[Fig.~2] The same as in Fig.~1, but in the $\tan\beta = 50$ case. 

\item[Fig.~3] The same as in Fig.~1, but for LEP2 energies.  

\item[Fig.~4] The same as in Fig.~3, but in the $\tan\beta = 50$ case. 

\end{description} 

\newpage 

\pagestyle{empty} 

%%%%%%%%%%%%%%%%%%%%%%%%%%%%%%%%%%%%%%%%%%%%%%%%%%%%%%%%%%%%%%%%
\begin{figure}[t]
\psfig{figure=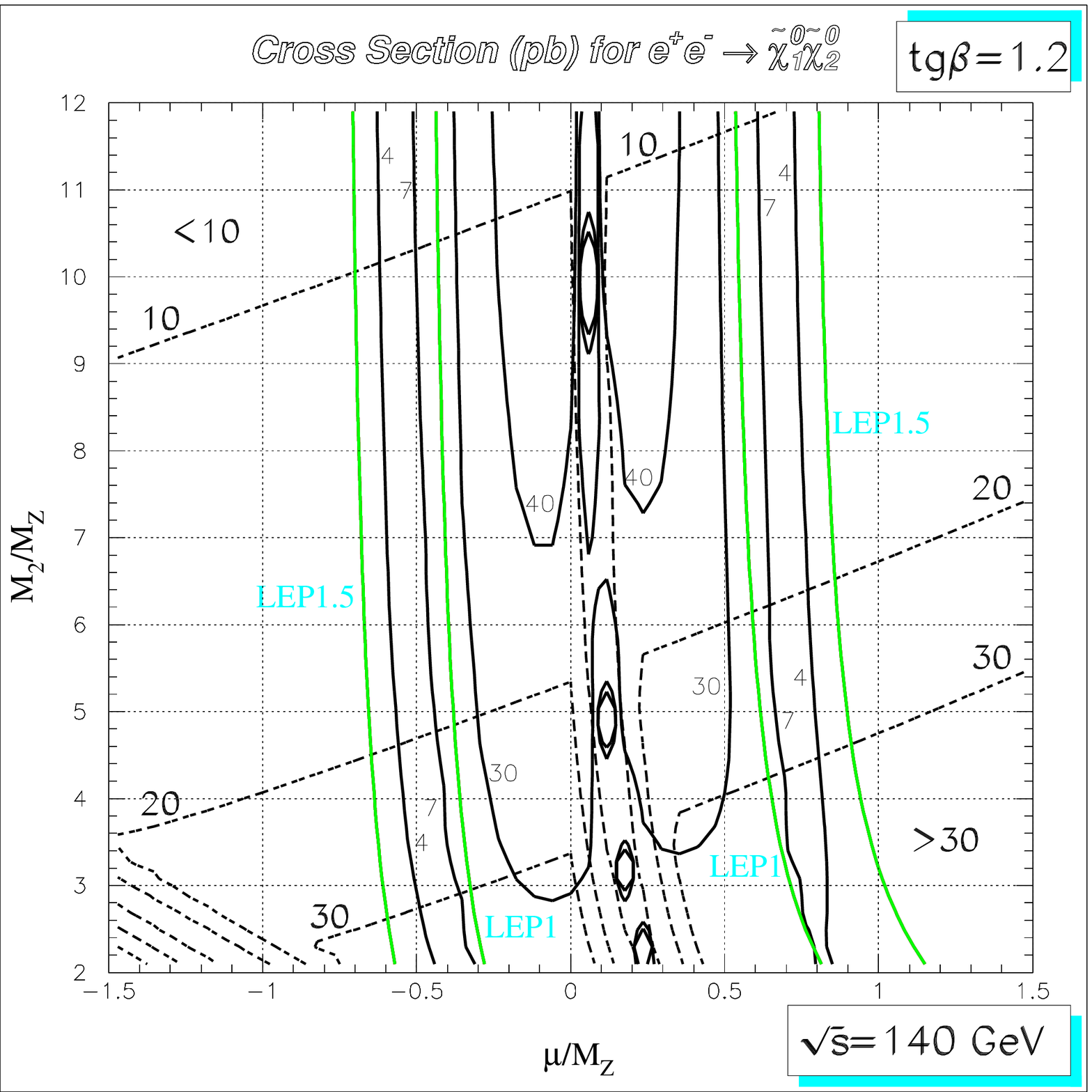,width=17.0cm,height=19.0cm}
\vspace{1.0truecm} 
\begin{center} 
{\Large \bf Figure 1a} 
\end{center} 
\end{figure}
%%%%%%%%%%%%%%%%%%%%%%%%%%%%%%%%%%%%%%%%%%%%%%%%%%%%%%%%%%%%%%%%

%%%%%%%%%%%%%%%%%%%%%%%%%%%%%%%%%%%%%%%%%%%%%%%%%%%%%%%%%%%%%%%%
\begin{figure}[t]
\psfig{figure=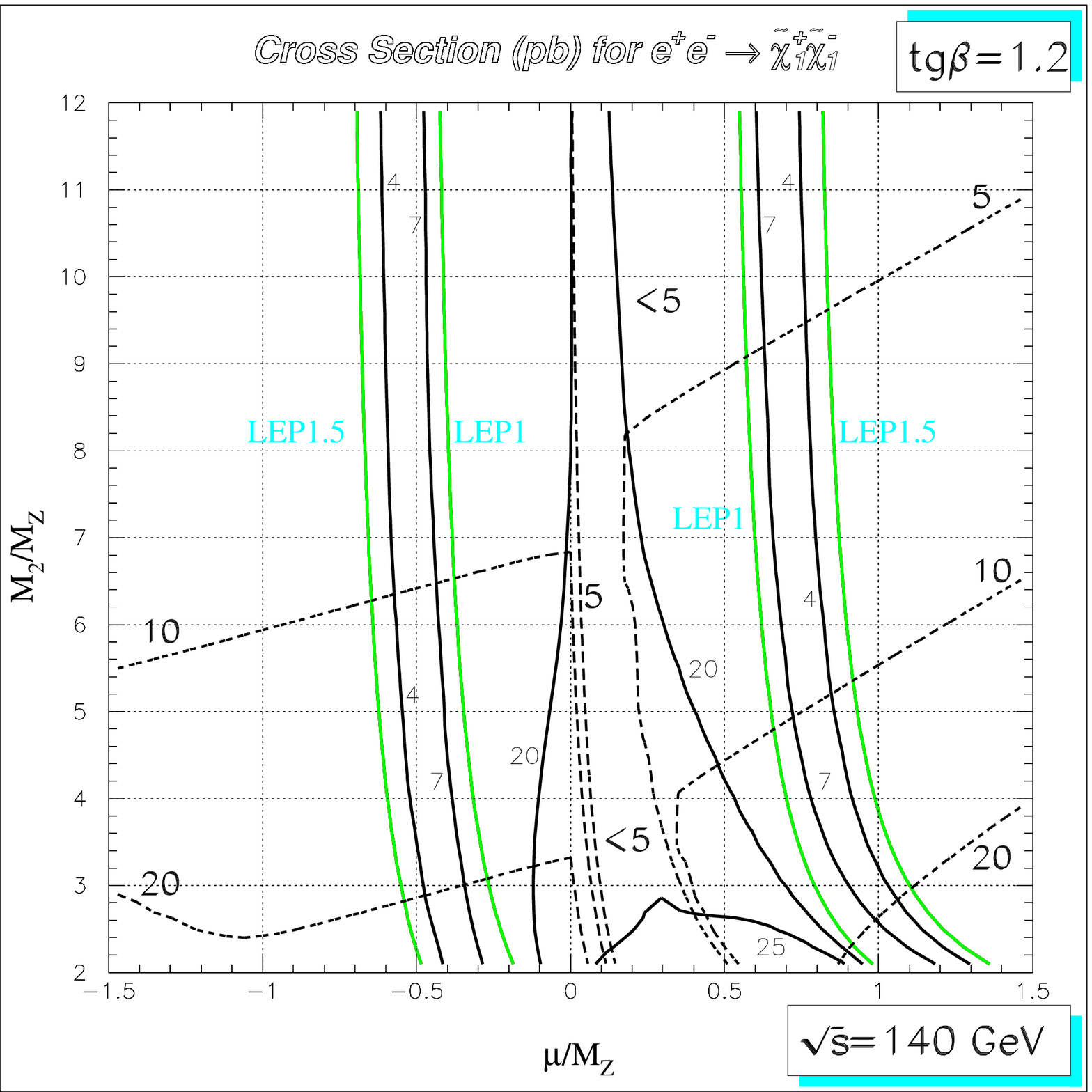,width=17.0cm,height=19.0cm}
\vspace{1.0truecm} 
\begin{center} 
{\Large \bf Figure 1b} 
\end{center} 
\end{figure}
%%%%%%%%%%%%%%%%%%%%%%%%%%%%%%%%%%%%%%%%%%%%%%%%%%%%%%%%%%%%%%%%

%%%%%%%%%%%%%%%%%%%%%%%%%%%%%%%%%%%%%%%%%%%%%%%%%%%%%%%%%%%%%%%%
\begin{figure}[t]
\psfig{figure=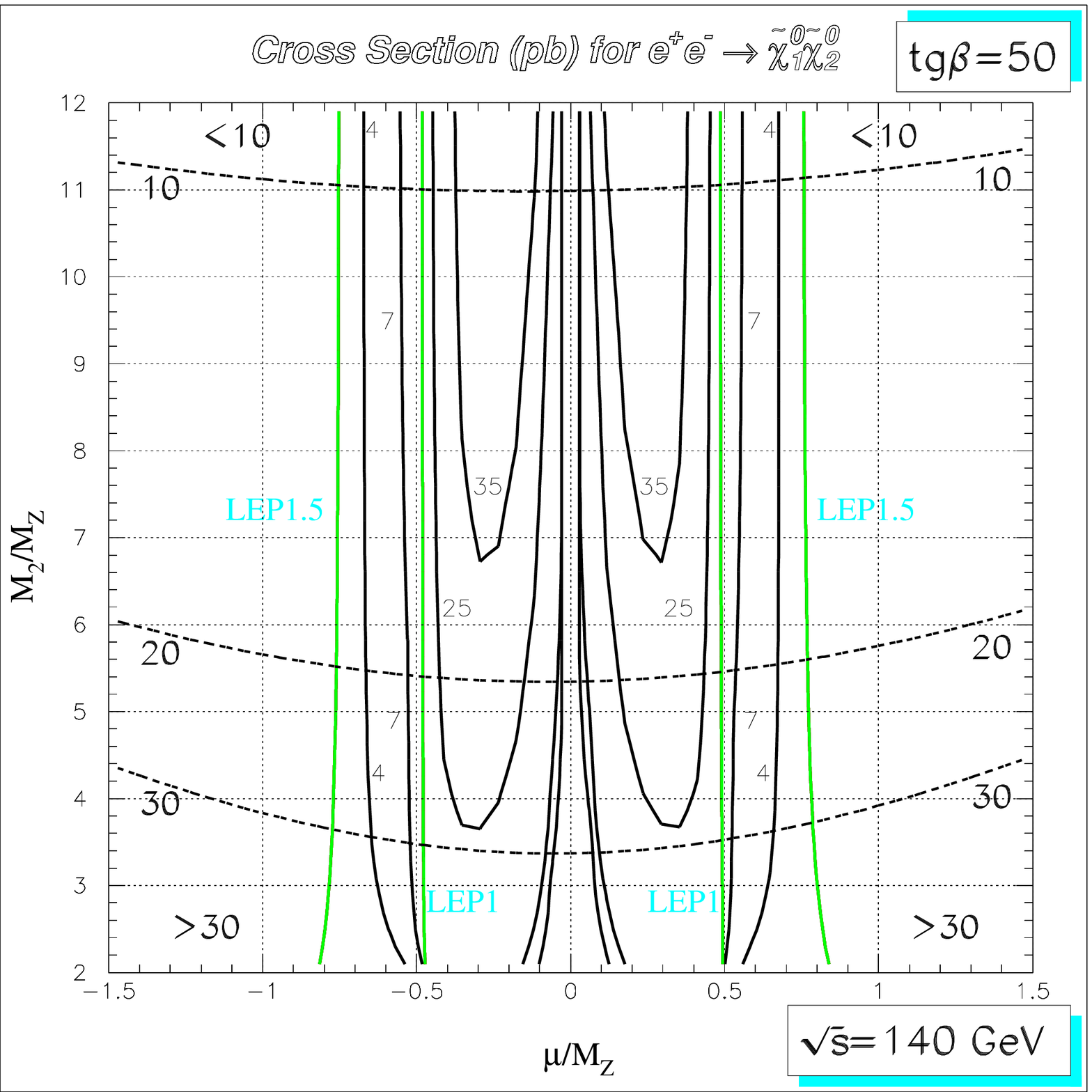,width=17.0cm,height=19.0cm}
\vspace{1.0truecm} 
\begin{center} 
{\Large \bf Figure 2a} 
\end{center} 
\end{figure}
%%%%%%%%%%%%%%%%%%%%%%%%%%%%%%%%%%%%%%%%%%%%%%%%%%%%%%%%%%%%%%%%

%%%%%%%%%%%%%%%%%%%%%%%%%%%%%%%%%%%%%%%%%%%%%%%%%%%%%%%%%%%%%%%%
\begin{figure}[t]
\psfig{figure=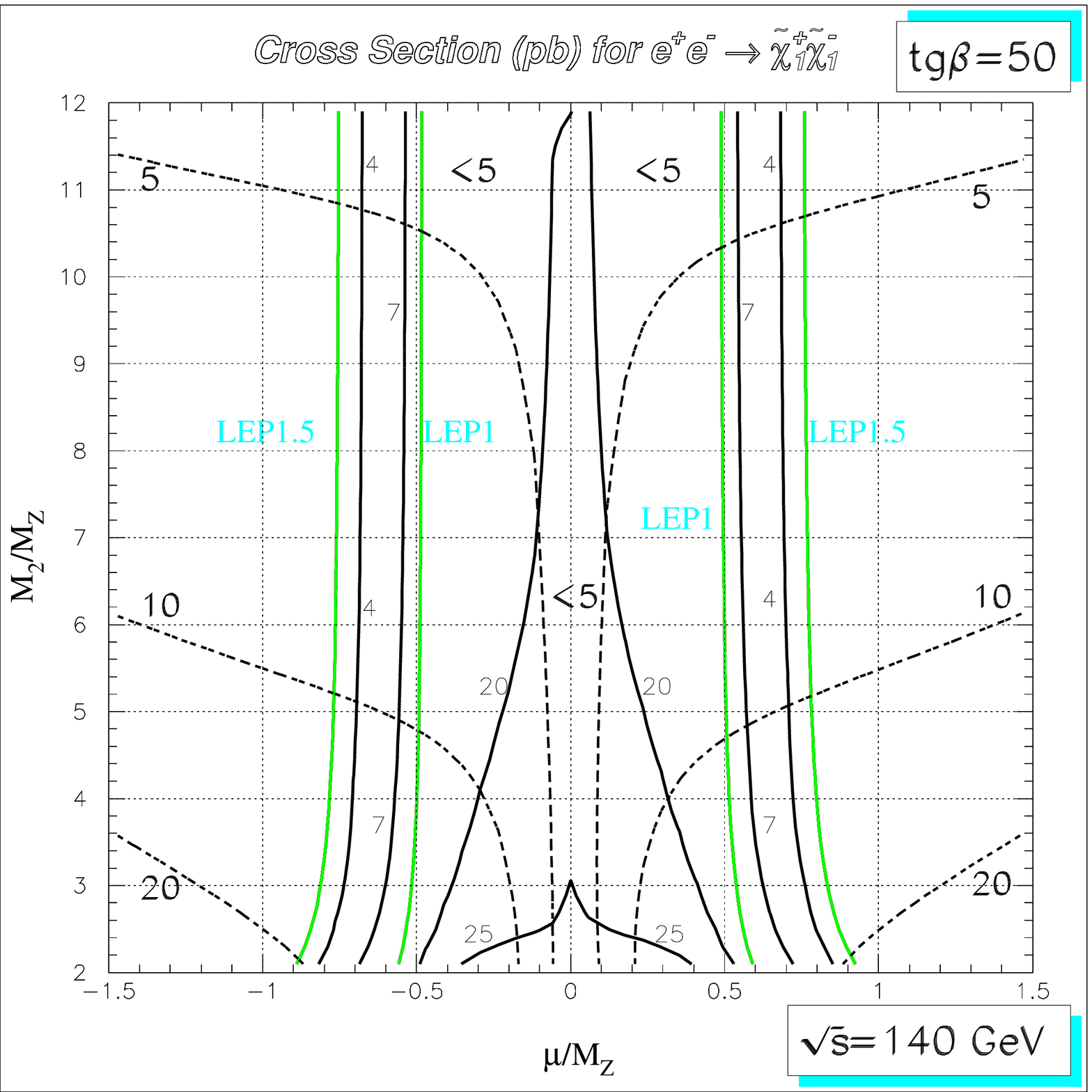,width=17.0cm,height=19.0cm}
\vspace{1.0truecm} 
\begin{center} 
{\Large \bf Figure 2b} 
\end{center} 
\end{figure}
%%%%%%%%%%%%%%%%%%%%%%%%%%%%%%%%%%%%%%%%%%%%%%%%%%%%%%%%%%%%%%%%

%%%%%%%%%%%%%%%%%%%%%%%%%%%%%%%%%%%%%%%%%%%%%%%%%%%%%%%%%%%%%%%%
\begin{figure}[t]
\psfig{figure=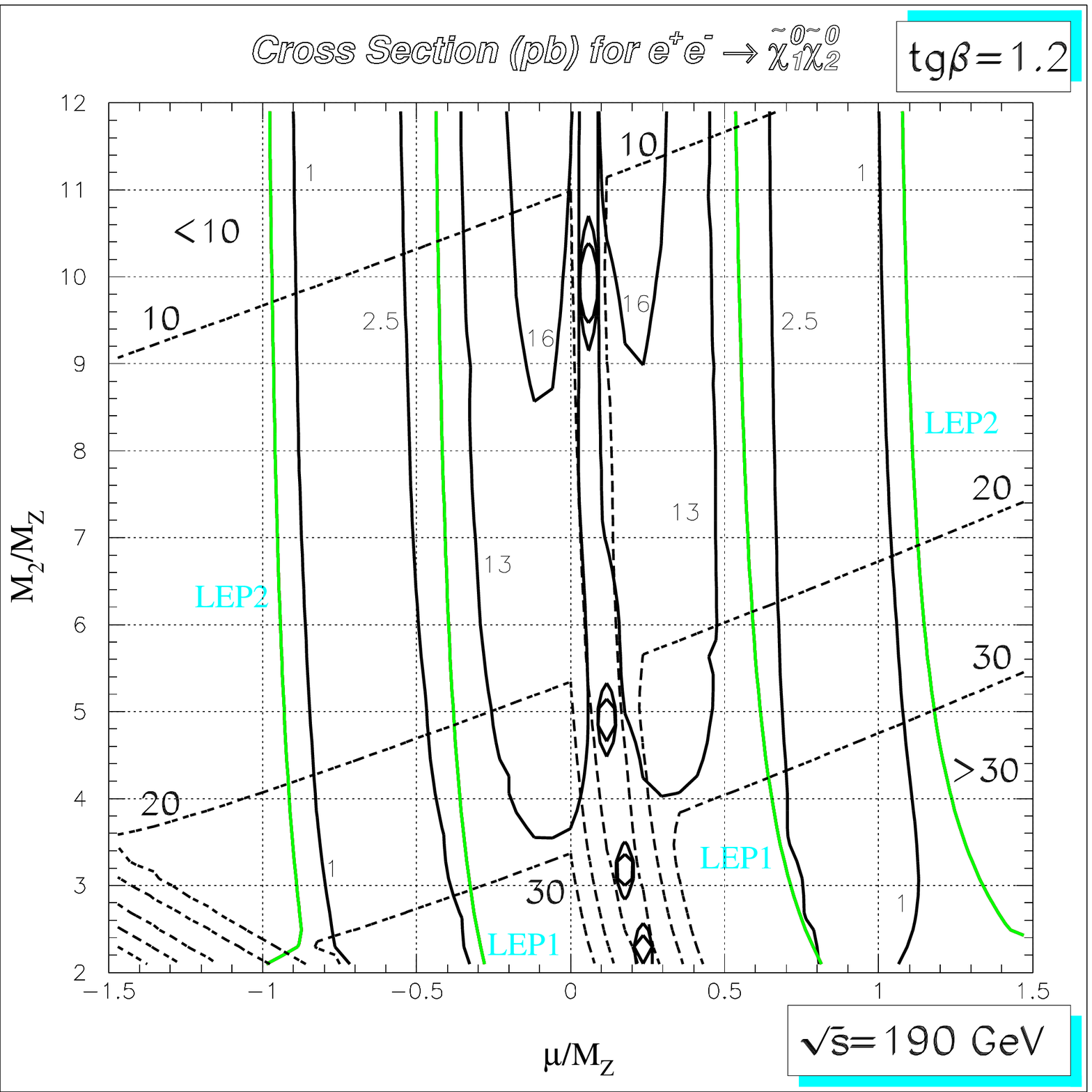,width=17.0cm,height=19.0cm}
\vspace{1.0truecm} 
\begin{center} 
{\Large \bf Figure 3a} 
\end{center} 
\end{figure}
%%%%%%%%%%%%%%%%%%%%%%%%%%%%%%%%%%%%%%%%%%%%%%%%%%%%%%%%%%%%%%%%

%%%%%%%%%%%%%%%%%%%%%%%%%%%%%%%%%%%%%%%%%%%%%%%%%%%%%%%%%%%%%%%%
\begin{figure}[t]
\psfig{figure=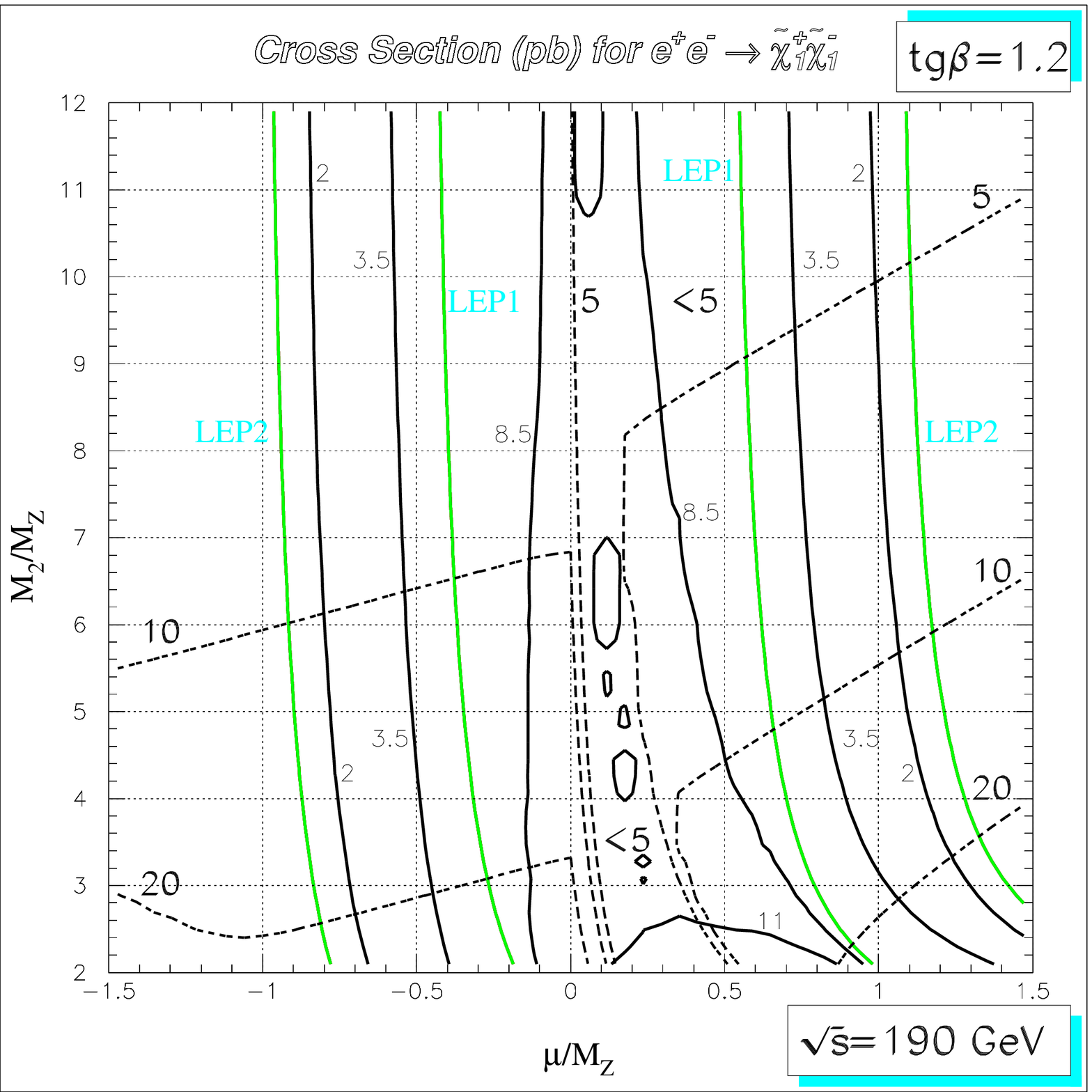,width=17.0cm,height=19.0cm}
\vspace{1.0truecm} 
\begin{center} 
{\Large \bf Figure 3b} 
\end{center} 
\end{figure}
%%%%%%%%%%%%%%%%%%%%%%%%%%%%%%%%%%%%%%%%%%%%%%%%%%%%%%%%%%%%%%%%

%%%%%%%%%%%%%%%%%%%%%%%%%%%%%%%%%%%%%%%%%%%%%%%%%%%%%%%%%%%%%%%%
\begin{figure}[t]
\psfig{figure=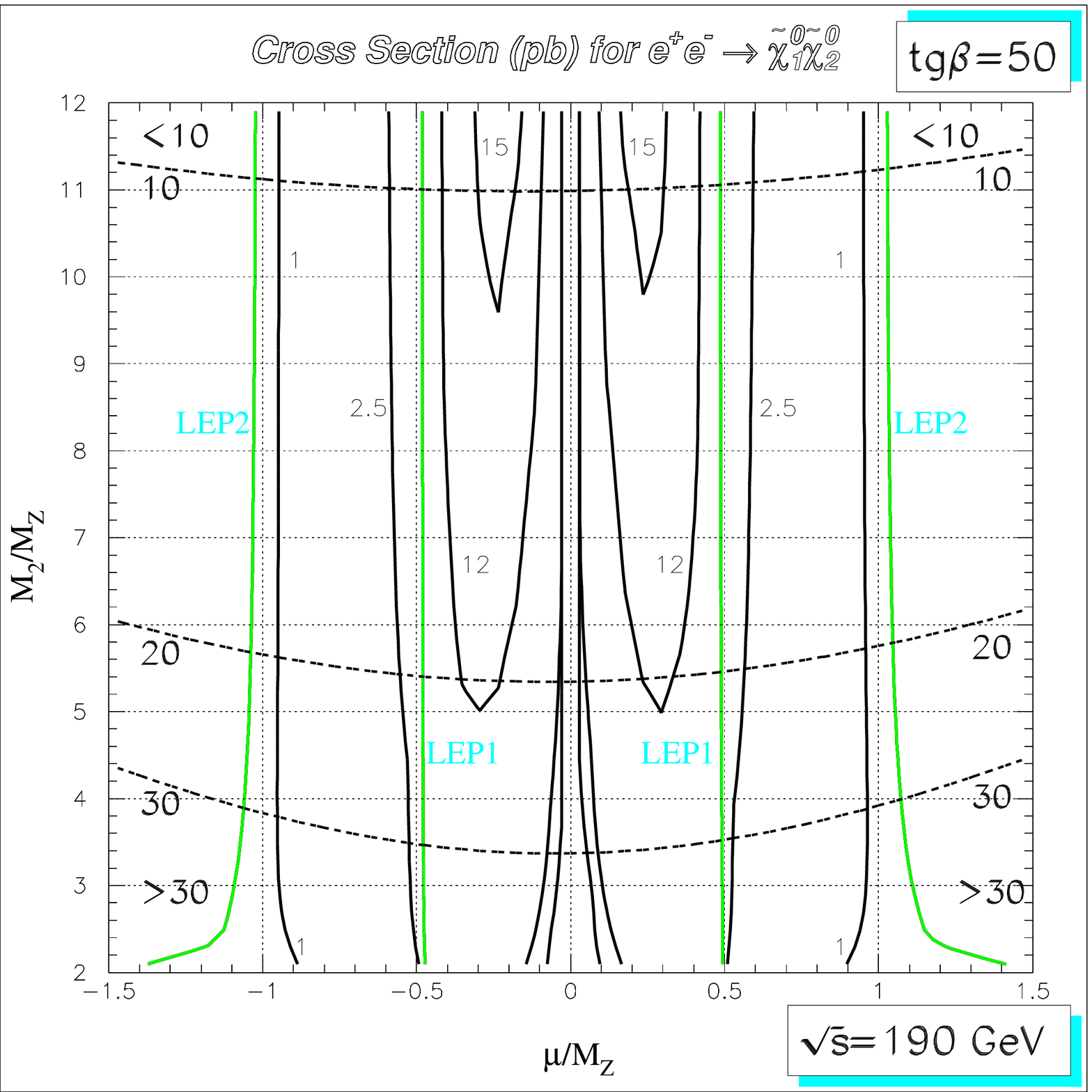,width=17.0cm,height=19.0cm}
\vspace{1.0truecm} 
\begin{center} 
{\Large \bf Figure 4a} 
\end{center} 
\end{figure}
%%%%%%%%%%%%%%%%%%%%%%%%%%%%%%%%%%%%%%%%%%%%%%%%%%%%%%%%%%%%%%%%

%%%%%%%%%%%%%%%%%%%%%%%%%%%%%%%%%%%%%%%%%%%%%%%%%%%%%%%%%%%%%%%%
\begin{figure}[t]
\psfig{figure=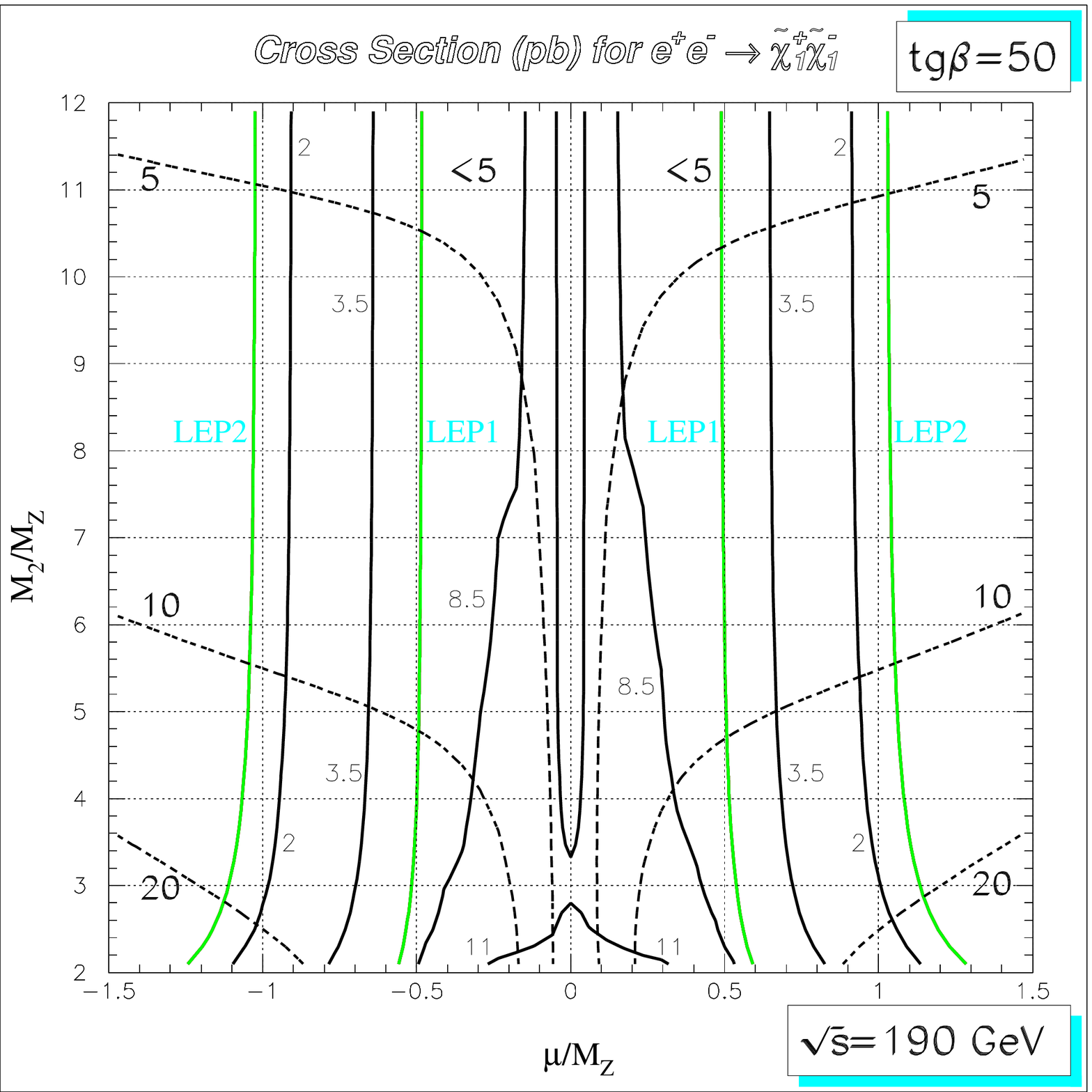,width=17.0cm,height=19.0cm}
\vspace{1.0truecm} 
\begin{center} 
{\Large \bf Figure 4b} 
\end{center} 
\end{figure}
%%%%%%%%%%%%%%%%%%%%%%%%%%%%%%%%%%%%%%%%%%%%%%%%%%%%%%%%%%%%%%%%


\begin{thebibliography}{99}

\bibitem{SUSYG} S.~Dimopoulos, S.~Raby and F.~Wilczek, \PRD{24}{81}{1681};\\
        S.~Dimopoulos and H.~Georgi, \NPB{193}{81}{150};\\
        L.~Iba\~nez and G.~G.~Ross, \PLB{105}{81}{150};\\
        H.P. Nilles, \PR{110}{84}{1}.

\bibitem{hk} H.~E.~Haber and G.~L.~Kane, \PR{117}{85}{75}. 

\bibitem{neumatr} J.~Ellis and G.~G.~Ross, \PLB{117}{82}{397}; \\ 
        J.~M.~Fr\`ere and G.~L.~Kane, \NPB{223}{83}{331}; \\ 
        J.~F.~Gunion and H.~E.~Haber, \NPB{272}{86}{1}; 
        {\bf B402} (1993) 567; \\ 
        A.~Bartl, H.~Fraas, W.~Majerotto and N.~Oshimo, \PRD{40}{89}{1594}. 

\bibitem{lepglobal} P.~Antilogus {\it et al.} [LEP Electroweak Working Group],
        LEPEWWG/95-02 (1995).

\bibitem{BF} M.~Boulware and D.~Finnel, \PRD{44}{91}{2054}.

\bibitem{ABF} G.~Altarelli, R.~Barbieri and F.~Caravaglios,
        \PLB{314}{93}{357}.

\bibitem{Gordy} J.~D.~Wells, C.~Kolda and G.~L.~Kane, \PLB{338}{94}{219}.

\bibitem{CW} M.~Carena and C.~E.~M.~Wagner, \NPB{452}{95}{45}. 

\bibitem{Joan} D.~Garcia and J.~Sola, \PLB{354}{95}{335}.

\bibitem{Stefan} P.~Chankowski and S.~Pokorski, to appear in 
        {\it Proc. Beyond the Standard Model IV}, Lake Tahoe, CA, 
        December 1994; MPI Preprint MPI-PhT/95-49. 

\bibitem{Hollik} A.~Dabelstein, W.~Hollik and W. ~M\"osle, 
        Univ. of Karlsruhe Preprint \\ KA-THEP-5-1995.

\bibitem{topquark} F.~Abe et al. [CDF Collaboration], \PRL{74}{95}{2626}; \\ 
        S.~Abachi et al. [D0 Collaboration], \PRL{74}{95}{2632}.

\bibitem{talk} C.~E.~M.~Wagner, talk presented at the {\it Conference 
        SUSY95}, Palaiseau, France, June 1995, to appear in the 
        {\it Proceedings}, CERN Preprint CERN-TH/95-261, 
        October 1995, hep-ph/9510341.

\bibitem{Yukawa} H. Arason {\it et al.}, \PRL{67}{91}{2933};\\  
         V.~Barger, M.~S.~Berger and P.~Ohmann,
        \PRL{49}{94}{4908}; \\
        P.~Langacker and N.~Polonsky, \PRD{47}{93}{4028}; 
        {\bf D49} (1994) 1454; \\
        M.~Carena, S.~Pokorski and C.~E.~M.~Wagner, \NPB{406}{93}{59}; \\
        W.~A.~Bardeen, M.~Carena, S.~Pokorski and C.~E.~M.~Wagner, 
        \PLB{320}{94}{110}.

\bibitem{largetb} M.~Olechowski and S.~Pokorski, \PLB{214}{88}{393};\\
        B.~Anantharayan, G.~Lazarides and Q.~Shafi, \PRD{44}{91}{1613};\\ 
        S.~Dimopoulos, L.~J.~Hall and S.~Raby, \PRL{68}{92}{1984}, 
        \PRD{45}{92}{4192}.

\bibitem{bartl} A.~Bartl, H.~Fraas and W.~Majerotto, \NPB{278}{86}{1}. 

\bibitem{Tata} D.~A.~Dicus and X.~Tata, \PRD{35}{87}{2110}; \\ 
        M. Chen, C.~Dionisi, M.~Martinez and X.~Tata, \PR{159}{88}{201}; \\ 
        J.~F.~Grivaz {\it et al.}, in {\it Proc. of the ECFA
        Workshop on LEP200}, Aachen, Sept. 29-Oct. 1, 1986, 
        eds. A. B\"ohm and
        W. Hoogland, CERN 87-08, ECFA 87/108, 
        Geneva, (1987) Vol. II, p. 380.

\bibitem{amb-mele} S.~Ambrosanio and B.~Mele, \PRD{52}{95}{3900}. 


\bibitem{amb-mele2}  S.~Ambrosanio and B.~Mele,
        ``Neutralino Decays in the Minimal Supersymmetric
        Standard Model'', Preprint ROME1-1095/95, August 1995,
        to be published in {\it Physical Review} {\bf D}, 
         hep-ph/9508237.

\bibitem{Bartl-ch} A.~Bartl, H.~Fraas and W.~Majerotto, \ZPC{30}{86}{441}; \\ 
        A.~Bartl, H.~Fraas, W.~Majerotto and 
        B.~M\"osslacher, \ZPC{55}{92}{257}.

\bibitem{interim} ``Interim Report on the Physics Motivations for an Energy 
Upgrade of LEP2", CERN Preprint CERN-TH/95-151, CERN-PPE/95-78.

\bibitem{dionisi} C. Dionisi, private communication.

\end{thebibliography}
\end{document}